# Energy efficiency analysis as a function of the working voltages in supercapacitors

**Jose J. Quintana[a], \*, Alejandro Ramos[a], Moises Diaz[b], Ignacio Nuez[a]**

[a] Universidad de Las Palmas de Gran Canaria, Las Palmas de Gran Canaria, Spain

[b] Universidad Del Atlantico Medio, Las Palmas de Gran Canaria, Spain

## Abstract

Supercapacitors are increasingly used as energy storage elements. Unlike batteries, their state of charge has a considerable influence on their voltage in normal operation, allowing them to work from zero to their maximum voltage. In this work, a theoretical and practical analysis is proposed of the energy efficiency of these devices according to their working voltages. To this end, several supercapacitors were subjected to charge and discharge cycles until the measurements of current and voltage stabilized. At this point their energy efficiency was calculated. These charge-discharge cycles were carried out: i) without rest between charging and discharging; and ii) with a rest of several minutes between the two stages. Using the information obtained from the tests, the energy efficiency is shown plotted against the minimum and maximum working voltages. By consulting the data and the graphs, the ideal working voltages to optimize the energy efficiency of these devices can be obtained.

## 1.- Introduction

Energy is a fundamental requirement to maintain the level of well-being that has been achieved by modern society. Currently, the primary source of energy comes from fossil fuels [1]. However, as is well-known, its growing use is exacerbating the greenhouse effect and, according to many studies, it is also driving climate change [2][3]. One way to mitigate this effect is to change from fossil fuels to renewable energies, such as wind and solar power. An example of their application can be seen in the electric sector. One of the main drawbacks of the greater integration of renewable energies is the problem of their storage [4][5][6][7][8]. While energy can be stored in many different ways, the use of electrical devices such as batteries and supercapacitors is gaining in popularity [9].

As the energy density of supercapacitors is approximately one-tenth that of batteries, they are not usually used as devices for storing energy [10]. However, the power density of supercapacitors is much higher than that of batteries, which makes them very useful in applications where the aim is to recover or deliver energy with high power peaks. A well-known example is the KERS (kinetic energy recovery system) [11] ¡, which stores the energy when a vehicle, such as a Formula 1 car, brakes and recycles it as the vehicle accelerates again. Another very important use is in lifts where energy savings of up to 70% can be made [12]. That is, they can be used in systems that do not need a considerable energy storage, but that can take advantage of energy peaks and then return them back to the system. This characteristic also makes them complementary to other storage systems[NO10][NO12] [13] [14] [15].

Another characteristic is that the temperature range of supercapacitors is also higher than that of batteries, and can range from -40 to 85°C. For this reason, they are widely used in starting systems for large engines and even for cars in very cold climates because at temperatures below 0°C the performance of batteries drops considerably and can render them inoperative. Supercapacitors, on the other hand, work at full capacity at temperatures down to -40°C and therefore guarantee that engines and cars will start at those temperatures [16].

Typically, the energy stored in a battery has very little influence on its voltage [17]. In contrast, the voltage on a supercapacitor is C times its charge, where C is its capacitance, and its stored energy is proportional to the square of that voltage. In ideal conditions, the supercapacitor could work from 0V to its maximum voltage [18]. However, a minimum working voltage is set in a real scenario as the currents required to supply a specific power may be very high at very low voltages.

Nevertheless, setting such a minimum voltage limits the use of the available energy in the device. For instance, if a supercapacitor works in a voltage range of 70% to 100% of its maximum voltage, it will only handle 50% of the device's maximum energy. Likewise, if it works from 50% to 100%, it will handle 75% of it. This implies a trade-off between the minimum working voltage and other factors such as energy availability and efficiency.

The literature on supercapacitors and the data sheets supplied by the manufacturers provide a great deal of information about the specific energy, specific power, and ageing of these devices [19]. However, very little has been published on the efficiency of these devices in relation to operating voltages. With this in mind, in this work, a study of supercapacitor energy efficiency is carried out considering different combinations of operating voltages. Firstly, a theoretical analysis of the efficiency is presented, which will subsequently be verified with experimental data.

This paper is organized as follows: Section 2 provides a brief review of supercapacitor efficiency calculation methods and gives the efficiency equations using charge-discharge cycles. Section 3 focuses on the tests that were carried out using various supercapacitors, while Section 4 discusses the results obtained. Finally, in section 5, the conclusions are drawn.

## 2.- Energy efficiency of supercapacitors

There are several procedures for calculating the characteristics of supercapacitors based on standards issued by different organizations [20] [21]. To calculate their efficiency as energy storage devices, [21] subjects the supercapacitor to charge-discharge cycles, while the IEC standard also includes rest times [20]. For this paper, approach [21] was chosen, although slightly modified.

The method proposed in this work is to subject the supercapacitor to constant current charge and discharge cycles. First, the current for the tests had to be defined. In [21], the current is calculated based on the capacitance of the supercapacitor, while in [20] the calculation is calculated based on the internal resistance and with the aim of achieving an efficiency of 95%. Because the objective in this paper is to compare efficiencies, the test currents were calculated based on [20] as they give a similar efficiency in all the devices. Since the temperature of the supercapacitor varied only very slightly, the stable state was defined when the electrical variables stabilized.

Once the current is defined, all the combinations of minimum and maximum voltages are defined for the tests. The supercapacitor is subjected to a certain number of charge-discharge cycles from minimum voltage to maximum voltage for each combination. This number of cycles will be defined in section 3.1. The test for each combination is performed three times on each supercapacitor. The data taken from each test are: the supplied and extracted current; the voltage at terminals; and the time. In this way, it is possible to work out the charge, the energy supplied and extracted in each cycle, and the energy efficiency of the supercapacitor.

Another factor to take into account is the quantification of current losses due to internal reorganization and leakage. In the tests [20], after reaching the maximum voltage, the supercapacitor is left charging at constant voltage for a time to partially neutralize the previous currents. In this work, our approach is quite different. Firstly, the efficiency is calculated in cycles in which there is no rest between charging and discharging, and secondly it is calculated when there is a rest period. By comparing these efficiencies it is possible to observe the effect of these phenomena.

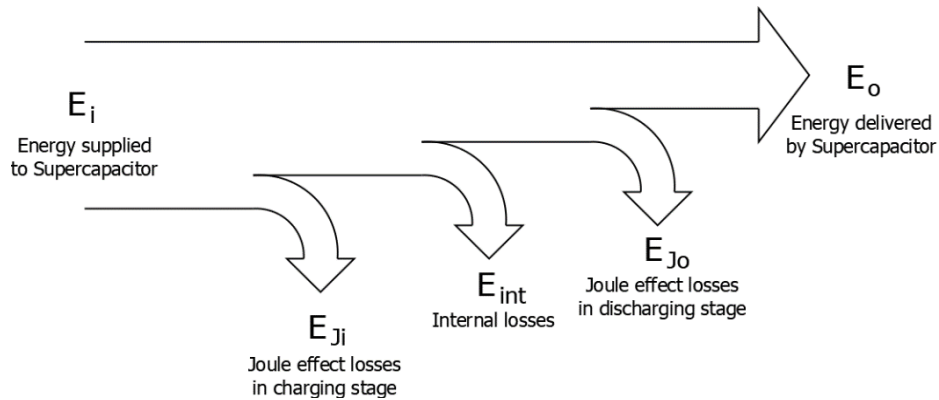

Figure 1.- Diagram of energies and losses in the supercapacitor.

The diagram in Figure 1 shows a visual representation of the energy efficiency calculations. The meanings of the symbols are as follows: $E_i$ denotes the energy supplied to the supercapacitor; $E_{Ji}$ is the energy dissipated in the charge stage because of the series resistance; $E_{int}$ is the energy lost due to having the energy stored for a specific time; $E_{Jo}$ is the energy lost because of the serial resistance when discharging; and $E_o$ is the energy supplied to an external load.

The energy efficiency, $\eta$, can be defined as the ratio between the energy supplied by the supercapacitor to an external load and the energy supplied to it in each cycle:

$$\eta = \frac{E_o}{E_i} \qquad (1)$$

The supercapacitor model of a resistor in series with an ideal capacitor [16] [21] is used to obtain the equation of energy efficiency. An ideal capacitor implies that both the charge and discharge curves at constant current are straight lines in the plots. To identify the value of the internal series resistance, the method is based on the voltage drop in the supercapacitor when the input current is abruptly cut [21]. The value of this resistance can then be calculated as a quotient between the voltage jump and the current.

## 2.1.- Energy efficiency in a cycle without rest

Figure 2 shows a real charge-discharge cycle in a supercapacitor. The beginning and end of the cycle are highlighted with circles. According to the basic model, the jumps ab, cd, de, and fa represent the voltage drop in the series resistor at charging and discharging. In this particular cycle, the charge stage starts at $t_{cs}$ and ends at $t_{ce}$. This process is carried out at a constant current, $i_c$, and ends when the supercapacitor reaches the maximum voltage set, $v_M$. The discharge stage starts at $t_{ds}$ and ends at $t_{de}$. It is also carried out at the same constant current, $i_c$, and ends when the supercapacitor reaches the minimum programmed voltage $v_m$.

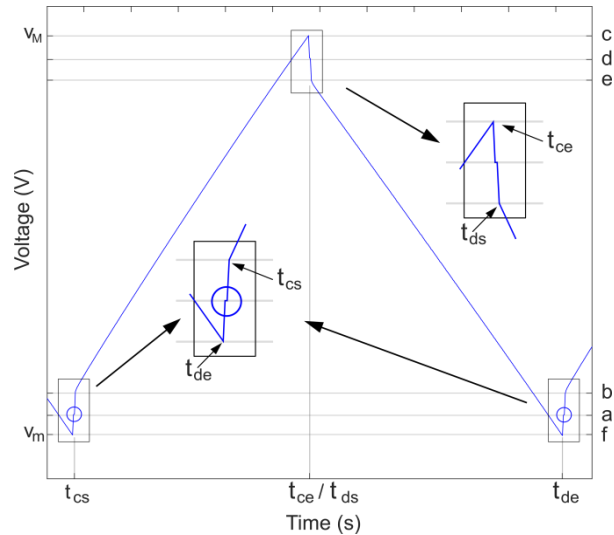

Figure 2.- Charge-discharge cycle without rest.

The energy efficiency can be calculated, in each cycle, as the relationship between the energy supplied to the supercapacitor, $E_i$, and the energy delivered to an external load, $E_o$. Assuming a general case where the charging and discharging currents are constant and equal, the energy efficiency is deduced as follows:

$$\eta = \frac{E_o}{E_i} = \frac{\int_{t_{ds}}^{t_{de}} v_o i_c dt}{\int_{t_{cs}}^{t_{ce}} v_i i_c dt} \quad (2)$$

To solve the integrals, the data and curves of Figure 2 are used. In each cycle there are four different periods, shown in (3):

$Period\ 1: t_{de} < t < t_{cs} \Rightarrow i = 0; v = v_a$

$Period\ 2: t_{cs} < t < t_{ce} \Rightarrow i = i_c$

$Period\ 3: t_{ce} < t < t_{ds} \Rightarrow i = 0; v = v_d$

$Period\ 4: t_{ds} < t < t_{de} \Rightarrow i = -i_c$

(3)

In periods 1 and 3, the energy transferred is zero. In period 2 the energy is supplied to the supercapacitor, and in period 4 the energy is extracted from it. To calculate the energy supplied in period 2, first we calculate the equation of the voltage and then the energy. As the behaviour of a supercapacitor is similar to that of a capacitor, the voltage on its terminals depends on the current, the capacitance and the internal resistance and, according to Figure 2, is:

$$v_i = v_a + i_c R + \int_{t_{cs}}^{t} \frac{i_c}{C} dt = v_a + i_c R + \frac{i_c}{C}(t - t_{cs}) \quad (4)$$

From Figure 2 it is observed that:

$$v_a = v_m + i_c R \quad (5)$$

Replacing the value of $v_a$ in equation (4), we have:

$$v_i = v_m + 2i_c R + \frac{i_c}{C}(t - t_{cs}) \quad (6)$$

From Figure 2, if $t = t_{ce}$ then $v_c = v_M$. Applied to equation (6), the following equality is obtained:

$$v_M = v_m + 2i_c R + \frac{i_c}{C}(t_{ce} - t_{cs}) \quad (7)$$

calculating the energy

$$E_i = \int_{t_{cs}}^{t_{ce}} v_i i_c dt$$

$$= i_c \int_{t_{cs}}^{t_{ce}} \left( v_m + 2i_c R + \frac{i_c}{C}(t - t_{cs}) \right) dt \quad (8)$$

$$= i_c \left[ (v_m + 2i_c R)(t_{ce} - t_{cs}) + \frac{i_c}{2C}(t_{ce}^2 - t_{cs}^2) - \frac{i_c}{C} t_{cs}(t_{ce} - t_{cs}) \right]$$

operating and rearranging

$$E_i = i_c \left( v_m + 2i_c R + \frac{i_c}{2C} t_{cs}(t_{ce} - t_{cs}) \right)(t_{ce} - t_{cs}) \quad (9)$$

combining this equation with (7) gives:

$$E_i = i_c \frac{v_M + v_m + 2i_c R}{2}(t_{ce} - t_{cs}) \qquad (10)$$

and operating in a similar way with the numerator:

$$E_o = \int_{t_{ds}}^{t_{de}} v_o i_c dt = i_c \frac{v_m + v_M - 2Ri_c}{2}(t_{de} - t_{ds}) \qquad (11)$$

gives the energy efficiency:

$$\eta = \frac{E_o}{E_i} = \frac{(v_m + v_M - 2Ri_c)(t_{de} - t_{ds})}{(v_m + v_M + 2Ri_c)(t_{ce} - t_{cs})} \qquad (12)$$

When the steady-state is reached in these devices, the charge supplied to the device at the charge stage is practically equal to the charge delivered at the discharge stage [21]. As a result, if the charging and discharging currents are equal, their respective times will also be equal, leaving the energy efficiency equation in the form:

$$\eta = \frac{v_M + v_m - 2i_c R}{v_M + v_m + 2i_c R} = 1 - \frac{4i_c R}{v_M + v_m + 2i_c R} \qquad (13)$$

It is observed that the factors that make the efficiency decrease are the working current and series resistance, and those that make it increase are the rise in the maximum and minimum working voltages.

## 2.2.- Energy efficiency in a cycle with rest

Figure 3 shows this type of cycle, with the beginning and end marked by circles. In this cycle, there are four stages. The first stage consists of charging at a constant current between points a and d; the second stage is the rest period between points d and e; the third is a discharge stage at a constant current between points e and h, and the final stage is another rest time between points h and a.

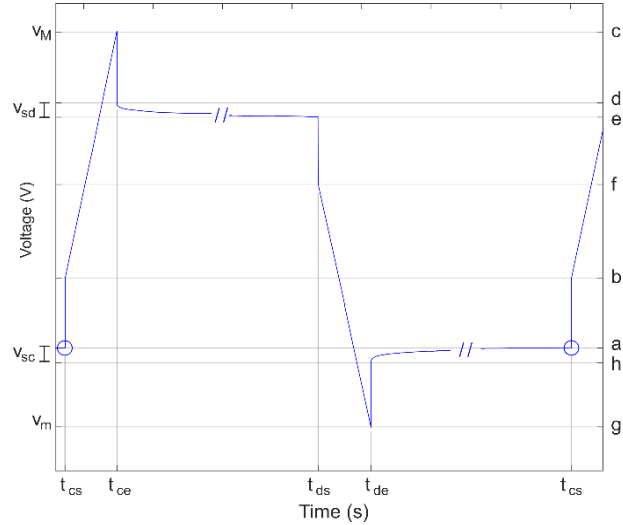

Figure 3.- Charge-discharge cycle with rest.

It should be noted that, when the supercapacitor's terminals are disconnected after the charge or discharge stage, the voltage measured in their terminals changes after a period of rest. Specifically, when the device is charged and its terminals are disconnected, the voltage drops. This phenomenon is known as self-discharge and is denoted as $v_{sd}$ [18]. Similarly, once the supercapacitor is discharged and its terminals are disconnected, the voltage at its terminals after a period of rest increases [22]. By analogy with self-discharge, in this paper we call this phenomenon self-charge and denote it as $v_{sc}$.

Taking into account the profile of the cycle in the steady-state shown in Figure 3, and operating as in the previous section, the energy supplied is given by:

$$E_i = \int_{t_{cs}}^{t_{ce}} v_i i_c dt = ic \int_{t_{cs}}^{t_{ce}} v_i dt = i_c \frac{v_m + v_M + 2Ri_c + v_{sc}}{2}(t_{ce} - t_{cs}) \tag{14}$$

and the energy delivered by:

$$E_o = \int_{t_{ds}}^{t_{de}} v_o i_c dt = i_c \frac{v_m + v_M - 2Ri_c - v_{sd}}{2}(t_{de} - t_{ds}) \tag{15}$$

Considering that, as in the previous case, the charge supplied and extracted in each charge-discharge cycle is practically the same, this implies that if the charge and discharge currents are equal, their times will also be equal. Therefore, operating as in the previous section, the efficiency of the supercapacitor is as follows:

$$\eta_R = \frac{E_o}{E_i} = \frac{v_M + v_m - v_{sd} - 2i_c R}{v_M + v_m + v_{sc} + 2i_c R} = 1 - \frac{4i_c R + v_{sc} + v_{sd}}{v_M + v_m + v_{sc} + 2i_c R} \tag{16}$$

It can be deduced that the efficiency increases with the working voltages and decreases as the device's internal resistance, the working current, and the self-discharge voltage increase.

# 3. Experimental data

Three 10F, 50F, and 100F Maxwell brand supercapacitors were all subjected to charge-discharge cycles. The charge and discharge currents for each supercapacitor were calculated in accordance with [20] as a function of their internal resistance. For the energy efficiency tests at variable voltages, these currents are 1.13A, 3.95A, and 4.7A for the 10F, 50F, and 100F capacitors, respectively.

As the supercapacitor has a maximum voltage of 2.7V, in the energy efficiency tests at different voltages all the voltages are referred to as this voltage in per-unit values. In this way, the minimum voltage is given by $V_m$ and the maximum by $V_M$. The voltages in per-unit values used for the tests were 0, 0.25, 0.50, 0.70, 0.90, and 1.0. Based on the different combinations of minimum and maximum voltages, the energy efficiency was calculated for each of the supercapacitors, both with and without rest. The tests were performed three times for each supercapacitor and the three supercapacitors were used for each capacitance. Therefore, nine samples were taken for each voltage combination and capacitance. The data used were the mean of these nine samples.

For data collection, in this work a programmable 10A current source was used to carry out the tests with the supercapacitors. The equipment allows the supercapacitor to be charged and discharged at constant current, leaving it with its terminals disconnected and short-circuited. Its simplified electrical circuit is shown in Figure 4. Data acquisition (current and voltage) was performed with a period of 0.1s and a resolution of ± 0.093 mV for voltage and ± 0.93 mA for current.

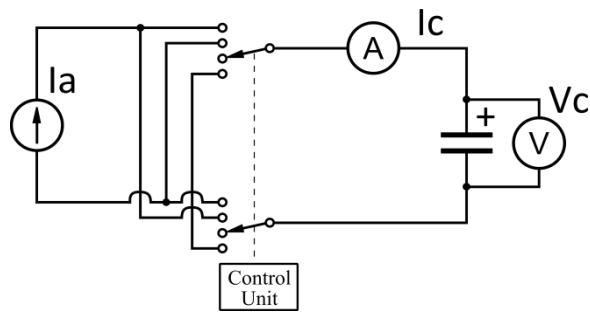

Figure 4.- Test equipment diagram.

## 3.1. Analysis of charge-discharge cycles

An analysis was undertaken of whether some parameters stabilize when the supercapacitor is subjected to charge-discharge cycles, both with and without rest. The parameters analysed were the cycle, charging and discharging times, the energy supplied and extracted, the charge supplied and extracted, and the energy efficiency. For this, the 10F supercapacitor was subjected to several charge-discharge cycles to analyse its behaviour and extrapolate it to the 50F and 100F supercapacitors. This test was performed at a current of 0.4A and between a minimum voltage of 0.5V and a maximum of 2.5V.

The first parameter analysed was the charge and discharge time in each period (Fig. 5). It can be seen that both times stabilize before the first ten cycles. It can be deduced from Figure 5 that the charge time is equal to the discharge time at the steady-state. Assuming that the charge current is equal to the discharge current, the charge supplied in the charge period is practically equal to the charge recovered in the discharge period. In [21], it is reported that when the difference between the amount supplied and extracted in a cycle is less than 1%, it can be assumed that the conditions are stable.

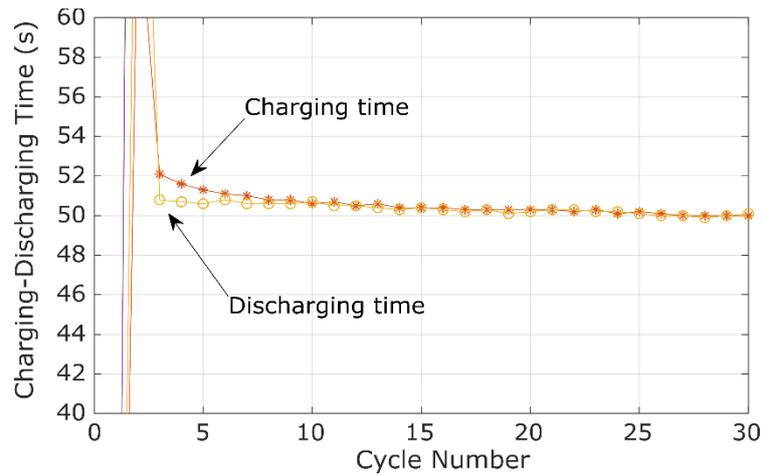

Figure 5.- Charging and discharging times in each cycle.

Another way of analyzing the relationship between the supplied charge, $Q_{in}$, and the extracted charge, $Q_{out}$, can be seen in Figure 6, which shows the ratio of these charges in each charge-discharge cycle with rest. Despite the rest, it can also be seen that there is almost no difference between the supplied and the extracted charge.

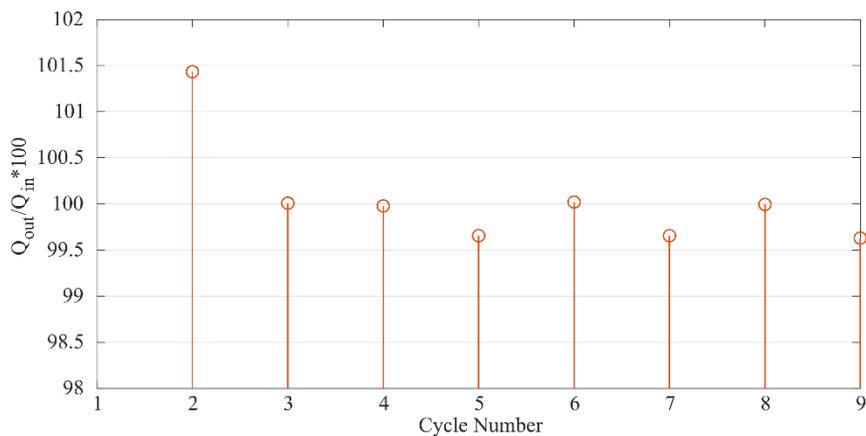

Figure 6.- Ratio of charge supplied and extracted in a supercapacitor in a charge-discharge cycle with a 10-minute rest time.

The following parameters analyzed were the energies in each cycle. Figure 7 shows the energy supplied to and recovered from the supercapacitor and the energy efficiency.

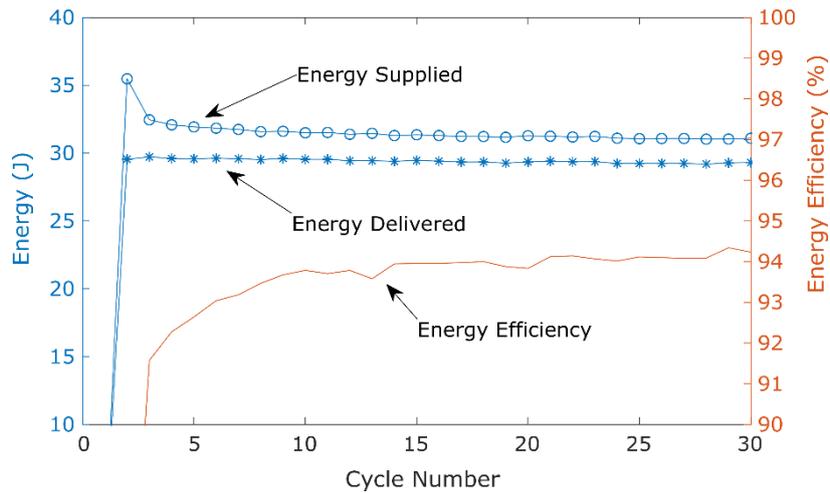

Figure 7.- Energy supplied, energy delivered and energy efficiency in each cycle.

As in the previous case, it can be seen that the parameters stabilize before the tenth cycle. It was therefore decided to use the mean values of the data collected between cycles 17 and 20 to carry out the energy efficiency studies for the calculations.

Equations (13) and (16) show that energy efficiency is related to the supercapacitor charging current, and that as it increases energy efficiency decreases. Table 1 shows experimental supercapacitor energy efficiencies at different charging currents when the minimum and maximum cycle voltages are 0.5V and 2.5V, respectively. It can be seen that the efficiency drops considerably when high currents are applied.

Table 1.- Energy efficiencies as a function of the current.

| 10F supercapacitor ||
|---|---|
| **Current (A)** | **Energy Efficiency (%)** |
| 0.4 | 95.2 |
| 0.5 | 94.5 |
| 0.75 | 92.9 |
| 2 | 84.1 |
| 4 | 76 |

## 3.2. Energy efficiency in cycles without rest

The energy efficiency η was calculated for each of the supercapacitors analyzed and for each pair of minimum and maximum voltages. The data are shown in Table 2.

Table 2.- Energy efficiency in cycles without rest.

| $V_m$ (P.U.) | $V_M$ (P.U.) | 10F η (%) | 50F η (%) | 100F η (%) |
|---|---|---|---|---|
| 0 | 0.25 | 56.1 | 63.5 | 62.3 |
| 0 | 0.5 | 75.0 | 79.3 | 79.0 |
| 0 | 0.7 | 81.0 | 84.3 | 84.9 |
| 0 | 0.9 | 84.3 | 87.2 | 87.8 |
| 0 | 1 | 85.4 | 88.1 | 89.1 |
| 0.25 | 0.5 | 83.7 | 86.0 | 85.3 |
| 0.25 | 0.7 | 86.3 | 88.5 | 88.6 |
| 0.25 | 0.9 | 88.0 | 90.1 | 90.5 |
| 0.25 | 1 | 88.5 | 90.7 | 91.2 |
| 0.5 | 0.7 | 89.4 | 91.2 | 90.5 |
| 0.5 | 0.9 | 90.4 | 92.2 | 92.1 |
| 0.5 | 1 | 90.7 | 92.3 | 92.4 |
| 0.7 | 0.9 | 91.4 | 92.9 | 92.8 |
| 0.7 | 1 | 91.8 | 93.4 | 93.3 |
| 0.9 | 1 | 94.4 | 94.1 | 94.1 |

To analyze these data, the energy efficiency of the supercapacitor was graphically represented as a function of the minimum and maximum voltages of the cycle. The minimum voltage is represented on the abscissa axis and the maximum on the ordinate axis in per-unit values. The energy efficiency values were calculated at the points marked x in the graphs, and with this information the plots of Figures 8 and 9 were generated. In Figure 8, the energy efficiency values are represented on the z-axis, while in Figure 9 the curves of equal energy efficiency are shown in a 2-D graph.

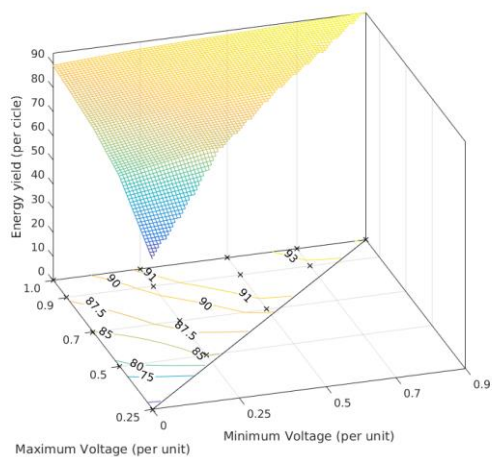

Figure 8.- Energy efficiency for 100F supercapacitor without rest.

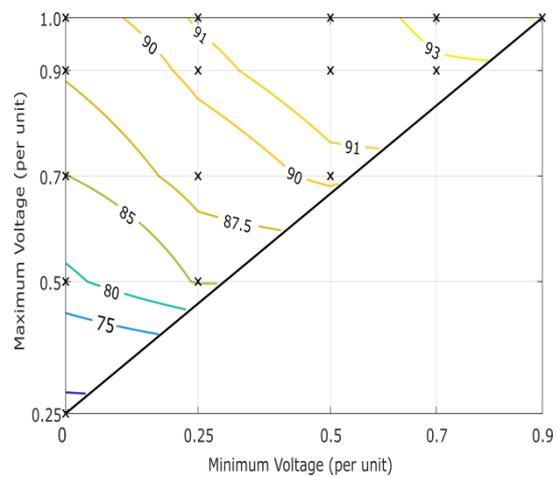

Figure 9.- Energy efficiency for 100F supercapacitor without rest, in equal energy efficiency curves.

For each minimum voltage, it can be seen that the maximum energy efficiency is obtained when the maximum voltage is the maximum voltage of the supercapacitor. It is also noted that energy efficiency increases when the maximum voltage is kept constant and the minimum voltage is increased. In contrast, energy efficiency decreases significantly when the supercapacitor is wholly discharged, i.e. when its minimum voltage is zero.

### 3.3. Energy efficiency in cycles with rest

This section describes the study that was undertaken of how the resting state influences the energy efficiency of the supercapacitor. For this purpose, the supercapacitors were subjected to charge-discharge cycles and were left with their terminals disconnected after the charge and discharge stage, as shown in Figure 3.

#### Influence of working voltages on resting voltage

It can be seen in Figure 3 that, with the terminals of the supercapacitor disconnected, their voltage does not remain constant but instead falls and rises after the charge and discharge stage, respectively. This behavior is due to its internal structure. The supercapacitor is formed by two porous electrodes made with materials of a large specific surface area. These electrodes are immersed in a solvent with an ionic liquid that allows formation of a double layer depending on the voltage at their terminals. The two electrodes are separated by a separator that galvanically isolates them but allows the ions to pass through [18]. When the voltage at the terminals of the supercapacitor is modified, there is a movement of ions inside it which does not stop immediately after leaving the terminals disconnected. This explains some of its properties, such as self-discharge [18] [23] [24] [25] and the loss of energy when the device has just been disconnected [26] [27].

For each test, the self-charge $V_{sc}$ and self-discharge $V_{sd}$ voltages were measured. A correlation was observed between these voltages and the difference between the maximum and minimum working voltages in the cycle. Table 3, which is ordered on the basis of this voltage difference, shows, for a supercapacitor of 50F and with a rest time of 30 minutes after charging and discharging, the self-charge, self-discharge and working voltages of the cycle.

Table 3.- Self-discharge and self-charge voltages as a function of cycle voltages in a 50F supercapacitor.

| $V_M-V_m$ (V) | $V_m$ (V) | $V_M$ (V) | Vsd (mV) | Vsc (mV) |
|---|---|---|---|---|
| 0.27 | 2.43 | 2.70 | 21 | 14 |
| 0.54 | 1.35 | 1.89 | 40 | 28 |
| 0.54 | 1.89 | 2.43 | 41 | 27 |
| 0.68 | 0.00 | 0.68 | 51 | 55 |
| 0.68 | 0.68 | 1.35 | 49 | 36 |
| 0.81 | 1.89 | 2.70 | 60 | 43 |
| 1.08 | 1.35 | 2.43 | 76 | 54 |
| 1.22 | 0.68 | 1.89 | 82 | 61 |
| 1.35 | 0.00 | 1.35 | 89 | 70 |
| 1.35 | 1.35 | 2.70 | 90 | 68 |
| 1.76 | 0.68 | 2.43 | 111 | 86 |
| 1.89 | 0.00 | 1.89 | 114 | 94 |
| 2.03 | 0.68 | 2.70 | 124 | 98 |
| 2.43 | 0.00 | 2.43 | 143 | 118 |
| 2.70 | 0.00 | 2.70 | 156 | 130 |

An analysis of the data shows that both the self-discharge and the self-charge voltages increase as the difference in cycle voltages increases. This behaviour was observed in all the supercapacitors analysed. Therefore, in the cycles with rest, it was expected, based on equation (16), that energy efficiency would decrease as the difference between the maximum and minimum voltage increased.

**Energy efficiency in cycles with rest**

As can be deduced from equation (16), the energy efficiency of supercapacitors decreases not only because of the resistance and the working current but also because of the self-charge and self-discharge voltages. According to which, the energy efficiency of these devices will decrease if energy is stored in them. With this in mind, the tests in the previous section were repeated with a rest time of 30 minutes after the charge and discharge stages. The experimental results are shown in Table 4.

Table 4.- Energy efficiency with a rest time of 30 minutes.

| $V_m$ (P.U.) | $V_M$ (P.U.) | 10F $\eta_R$ (%) | 50F $\eta_R$ (%) | 100F $\eta_R$ (%) |
|---|---|---|---|---|
| 0 | 0.25 | 44.3 | 55.0 | 56.9 |
| 0 | 0.5 | 59.3 | 69.6 | 72.0 |
| 0 | 0.7 | 64.5 | 74.6 | 77.4 |
| 0 | 0.9 | 67.5 | 77.5 | 80.3 |
| 0 | 1 | 68.5 | 78.3 | 81.3 |
| 0.25 | 0.5 | 76.6 | 82.2 | 83.2 |
| 0.25 | 0.7 | 76.9 | 83.1 | 84.5 |
| 0.25 | 0.9 | 77.1 | 83.9 | 85.5 |
| 0.25 | 1 | 77.0 | 84.1 | 85.9 |
| 0.5 | 0.7 | 86.0 | 88.6 | 89.2 |
| 0.5 | 0.9 | 84.3 | 88.4 | 89.4 |
| 0.5 | 1 | 83.6 | 88.1 | 89.3 |
| 0.7 | 0.9 | 88.5 | 90.4 | 91.3 |
| 0.7 | 1 | 87.9 | 90.7 | 91.5 |
| 0.9 | 1 | 89.0 | 91.5 | 93.0 |

As in the previous section, the energy efficiency of the supercapacitor is graphically represented as a function of the minimum and maximum voltages of the cycle.

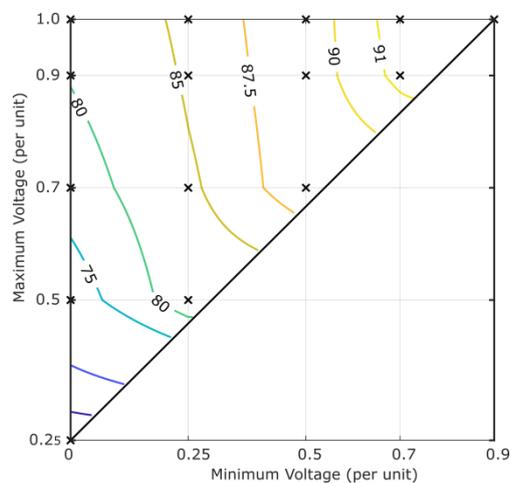

Figure 10.- Energy efficiency in 100F supercapacitor with rest.

It is observed, in comparison with the cycles without rest that, in all combinations and for the same supercapacitor, energy efficiency is lower when energy is stored during a time, as was deduced in equation (16).

# 4. Discussion

Equations (13) and (16) and the experimental data show that the energy efficiency of a supercapacitor is not constant and depends on several factors. One of the most important of these is series resistance. This parameter varies depending on the supercapacitor in question and is usually given in the datasheets provided by the manufacturer. As a general rule, a supercapacitor with as low a series resistance as possible should be chosen.

Another critical factor is the working current. These devices have a nominal current that is primarily a function of their internal resistance and is typically calculated for the supercapacitor to provide 95% energy efficiency [20]. Since these devices usually work with currents much higher than their nominal current, it is important to note that in such a circumstance their energy efficiency will decrease considerably, as was shown in section 3.2.

As for the working voltages, the experimental data and the generated curves show that, if the minimum voltage is fixed, the maximum energy efficiency is obtained when the supercapacitor works at its maximum voltage.

It was also found that energy efficiency decreases when the maximum voltage is fixed and the minimum voltage is decreased, with energy efficiency dropping appreciably when the supercapacitor is fully discharged, i.e. when its minimum voltage is zero. This contrasts with the case of available energy, which is higher the lower the minimum operating voltage, and so a compromise between energy efficiency and available energy must be reached.

On the other hand, when the supercapacitor is used to store energy for long periods, its energy efficiency is lower than when the energy is used immediately. It was observed that the self-discharge process also lowers energy efficiency and that this effect is more significant as the difference between minimum and maximum working voltage increases.

# 5. Conclusion

In this work, an analysis was undertaken of the energy efficiency of different supercapacitors used as energy storage devices according to their working voltages. This analysis was carried out in two situations: when the supercapacitor energy is used immediately and when it is used after a rest period. The procedure followed to measure energy efficiency in each voltage combination was to subject the supercapacitors to charge-discharge cycles until their measurements stabilized.

Using the values of the steady-state cycle, an equation for energy efficiency was deduced which was a function of the maximum and minimum working voltages, the working current, and the internal resistance. In addition, in the cycles with rest, energy efficiency also depended on the self-discharge and self-charge voltages.

The experimental data and the graphs confirm the deductions that were made from the energy efficiency equations that were obtained, namely that the energy efficiency of the supercapacitors decreases with increasing working current, and increases with decreasing internal resistance . Regarding the working voltages, in cycles without rest, it was observed that if the supercapacitor works between 0.7 and the maximum voltage its energy efficiency is above 93%, while if it works between 0.5 and 1 it never drops below 90%. On the other hand, if there is an energy storage period of 30 minutes, in the first case energy efficiency is greater than 91%, and in the second it is greater than 88%. It was thus observed that energy storage reduces the energy efficiency of the device.

Regarding the relationship between energy efficiency and working voltages, energy efficiency was represented as a function of the minimum and maximum working voltages. Firstly, it was observed that, in order to obtain the maximum energy efficiency from these devices, the maximum operating voltage must be the maximum supported by the device. Secondly, the device should not be fully discharged as its energy efficiency drops dramatically. Thirdly, when the minimum working voltage increases, so too does the energy efficiency, but at the cost of using less of its stored energy.

## Acknowledgements

Support of grant ULPGC2014-08 "Modelado Y Experimentación De Los Supercondensadores Como Dispositivo Para Almacenamiento Energético" is gratefully acknowlegded.

## Bibliography

[1] Shaping a secure and sustainable energy future for all. Int Energy Agency 2020. http://www.iea.org/ (accessed September 22, 2020).

[2] Creutzig F, Roy J, Lamb WF, Azevedo IML, Bruine de Bruin W, Dalkmann H, et al. Towards demand-side solutions for mitigating climate change. Nat Clim Chang 2018;8:260–3. https://doi.org/10.1038/s41558-018-0121-1.

[3] Shugar DH, Burr A, Haritashya UK, Kargel JS, Watson CS, Kennedy MC, et al. Rapid worldwide growth of glacial lakes since 1990. Nat Clim Chang 2020. https://doi.org/10.1038/s41558-020-0855-4.

[4] Fact Sheet: Energy Storage (2019) | White Papers | EESI n.d. https://www.eesi.org/papers/view/energy-storage-2019 (accessed February 24, 2020).

[5] Zhang F, Zhao P, Niu M, Maddy J. The survey of key technologies in hydrogen energy storage. Int J Hydrogen Energy 2016;41:14535–52. https://doi.org/10.1016/j.ijhydene.2016.05.293.

[6] Beaudin M, Zareipour H, Schellenberglabe A, Rosehart W. Energy for Sustainable Development Energy storage for mitigating the variability of renewable electricity sources : An updated review. Energy Sustain Dev 2010;14:302–14. https://doi.org/10.1016/j.esd.2010.09.007.

[7] Meng X, Shi L, Yao L, Zhang Y, Cui L. Novel Designs of Hybrid Thermal Energy Storage System and Operation Strategies for Concentrated Solar Power Plant. Energy

2020:124658. https://doi.org/10.1016/j.energy.2020.119281.

[8]  Olabi AG, Onumaegbu C, Wilberforce T, Ramadan M, Abdelkareem MA, Al – Alami AH. Critical Review of Energy Storage Systems. Energy 2020:118987. https://doi.org/10.1016/j.energy.2020.118987.

[9]  Mamen A, Supatti U. A survey of hybrid energy storage systems applied for intermittent renewable energy systems. ECTI-CON 2017 - 2017 14th Int Conf Electr Eng Comput Telecommun Inf Technol 2017:729–32. https://doi.org/10.1109/ECTICon.2017.8096342.

[10] Inthamoussou FA, Pegueroles-Queralt J, Bianchi FD. Control of a supercapacitor energy storage system for microgrid applications. IEEE Trans Energy Convers 2013;28:690–7. https://doi.org/10.1109/TEC.2013.2260752.

[11] Pipitone E, Vitale G. A regenerative braking system for internal combustion engine vehicles using supercapacitors as energy storage elements - Part 1: System analysis and modelling. J Power Sources 2020;448:227368. https://doi.org/10.1016/j.jpowsour.2019.227368.

[12] Skeleton Technologies - Industrial applications n.d. https://www.skeletontech.com/industrial-applications (accessed March 5, 2021).

[13] Aktaş A, Kırçiçek Y. A novel optimal energy management strategy for offshore wind/marine current/battery/ultracapacitor hybrid renewable energy system. Energy 2020;199. https://doi.org/10.1016/j.energy.2020.117425.

[14] Wang Y, Sun Z, Li X, Yang X, Chen Z. A comparative study of power allocation strategies used in fuel cell and ultracapacitor hybrid systems. Energy 2019;189:116142. https://doi.org/10.1016/j.energy.2019.116142.

[15] Peng H, Wang J, Shen W, Shi D, Huang Y. Compound control for energy management of the hybrid ultracapacitor-battery electric drive systems. Energy 2019;175:309–19. https://doi.org/10.1016/j.energy.2019.03.088.

[16] Skeleton Technologies - Engine start module n.d. https://www.skeletontech.com/skelstart-ultracapacitor-engine-start-module (accessed March 5, 2021).

[17] Saldaña G, Martín JIS, Zamora I, Asensio FJ, Oñederra O. Analysis of the current electric battery models for electric vehicle simulation. Energies 2019;12. https://doi.org/10.3390/en12142750.

[18] Conway BE. Electrochemical supercapacitors: Scientific fundamentals and technological applications. New York: Kluwer Academic/Plenum; 1999.

[19] Supercapacitor M, Note T. Murata Supercapacitor Technical Note n.d.:1–39.

[20] Standard IEC 62576:2009, Electric double-layer capacitors for use in hybrid electric vehicles 2009.

[21] FreedomCAR Ultracapacitor Test Manual. US Dep Energy 2004:DOE/NE-ID-11173.

[22] Mundy A, Plett GL. Reduced-order physics-based modeling and experimental parameter identification for non-Faradaic electrical double-layer capacitors. J Energy

Storage 2016;7:167–80. https://doi.org/10.1016/j.est.2016.06.009.

[23] Kaus M, Kowal J, Sauer DU. Modelling the effects of charge redistribution during self-discharge of supercapacitors. Electrochim Acta 2010;55:7516–23. https://doi.org/10.1016/j.electacta.2010.01.002.

[24] Zubieta L, Bonert R. Characterization of double-layer capacitors for power electronics applications. IEEE Trans Ind Appl 2000;36:199–205. https://doi.org/10.1109/28.821816.

[25] Lewandowski A, Jakobczyk P, Galinski M, Biegun M. Self-discharge of electrochemical double layer capacitors. Phys Chem Chem Phys 2013;15:8692–9. https://doi.org/10.1039/c3cp44612c.

[26] Nuez I, Quintana JJ, Ortega J. Modelo termico de los supercondensadores EDLC para aplicaciones en almacenamiento de energia. Rev La Acad Canar Ciencias 2014;26:61–81.

[27] Tao H, Lian C, Liu H. Multiscale modeling of electrolytes in porous electrode: From equilibrium structure to non-equilibrium transport. Green Energy Environ 2020;5:303–21. https://doi.org/10.1016/j.gee.2020.06.020.